\documentstyle[twocolumn,epsfig,prl,aps]{revtex}
\begin{document}
\draft
\title{ Observation of Andreev Surface Bound States in the 3-K Phase region of Sr$_2$RuO$_4$}
\twocolumn[
\hsize\textwidth\columnwidth\hsize\csname@twocolumnfalse\endcsname
\author{Z. Q. Mao, K. D. Nelson, R. Jin, and Y. Liu}
\address{Department of Physics, The Pennsylvania State University,
University Park, PA 16802}
\author{Y. Maeno}
\address{Department of Physics, Kyoto University, Kyoto 606-8502, Japan and\\
Core Research for Evolutional Science and Technology (CREST), Japan Science and Technology Corporation, Kawaguchi, Saitama 332-0012, Japan}
\date{\today}
\maketitle

\begin{abstract}
The tunneling spectrum of the superconducting phase with $T_{\rm c}\simeq 3.0$ K has been measured in the Ru-embedded region of Sr$_2$RuO$_4$ using cleaved junctions. A sharp zero-bias conductance peak (ZBCP) has been observed below 3 K. All characteristics of this ZBCP suggest that it originates from Andreev surface bound states, indicating that the pairing in the 3-K phase is also non-$s$-wave. Below the bulk $T_{\rm c}$ of Sr$_2$RuO$_4$ ($\simeq 1.5$ K), a bell-shaped ZBCP was found. This supports that there is a phase transition in the 3-K phase region near the bulk $T_{\rm c}$. 
\end{abstract}

\pacs{74.50.+r,74.25.Fy,74.70.-b}
]
\narrowtext

Several recent experiments \cite{1,2,3,4,5,6} have firmly established the non-$s$-wave nature of superconductivity in the layered perovskite Sr$_2$RuO$_4$ \cite{7}. In particular, the NMR experiment \cite{3} has provided strong evidence that bulk Sr$_2$RuO$_4$ is a spin-triplet superconductor, as theoretically anticipated \cite{8}. The intrinsic superconducting critical temperature ($T_{\rm c0}$) of Sr$_2$RuO$_4$ is 1.5 K \cite{2,9}. While searching for the optimal crystal growth conditions, an eutectic solidification system, Ru metal embedded in the primary phase of Sr$_2$RuO$_4$, was found. An intriguing observation was that the $T_{\rm c}$ of this eutectic system was enhanced up to 3 K \cite{10}. It has been suggested that the enhanced superconductivity at 3 K most probably originate at the interface region between the Ru inclusion and Sr$_2$RuO$_4$ single crystal \cite{10,11}.
 
However, the nature of the 3-K phase, as well as the origin of the $T_{\rm c}$ enhancement by Ru embedding, has not been clarified. Addressing these issues is certainly important, as it may provide new insight into the mechanism leading to $p$-wave superconductivity in bulk Sr$_2$RuO$_4$. The experimental study for this phase is difficult to carry out using conventional techniques such as specific heat and NMR due to its small volume fraction. However, tunneling spectroscopy measurements, a powerful probe for the study of superconductivity, is not subject to this limitation. If the 3-K phase is unconventional, a zero-bias conductance peak (ZBCP) is expected in the tunneling spectrum as in the case of high-$T_c$ cuprates \cite{12,13,14,15,16} and that of heavy-fermion superconductors \cite{17}. This ZBCP is a result of the surface low-energy mid-gap states formed via Andreev reflection (also referred to as surface Andreev bound states (ABS)) \cite{12}, which can only exist when the phase of the pairing amplitude depends on the direction.

In this Letter, we report results of our tunneling measurements on $c$-axis junctions prepared by cleaving Ru-containing Sr$_2$RuO$_4$ single crystals along the $ab$ plane. We observed a ZBCP below 3K, attributable to ABSs, indicating that the 3-K phase is also non-$s$-wave.

The single crystals of Sr$_2$RuO$_4$ were grown by a floating-zone technique \cite {9}. The onset $T_{\rm c}$ measured by AC susceptibility $\chi_{\rm ac}$ was 1.40 K. Careful examination by an optical microscope revealed a few scattered Ru inclusions in parts of the crystal of this batch. The Ru inclusions are of needle-like shape with diameter of 1 $\mu$m and length of 10-30 $\mu$m. 

Several $c$-axis cleaved junctions were prepared from this batch of crystals. The cleaved crystals were fixed tightly by a Teflon frame with a thermal contraction coefficient much larger than Sr$_2$RuO$_4$, or by an insulated metal frame with a set screw. Among these junctions, we observed two generic types of tunneling spectrum of d$I$/d$V$: One showing a ZBCP associated with the 3-K superconducting phase (reproducibly observed in three junctions), and the other a dip feature at zero bias unrelated to the superconductivity. From our analysis given below, it can be seen that Ru-inclusions are involved at the cleaved interfaces in the former, denoted by SRO/Ru/SRO below. The data presented below for the SRO/Ru/SRO junction were obtained on the same sample.

All current-voltage ($I-V$) characteristics, as well junction resistance $R_{\rm J}(T)$, were measured by a four-point probe in a $^3$He cryostat. To reduce heating, the $I-V$ curve was measured by d.c. pulsed-current method with a typical duration of 50 ms followed by a 2s delay between two successive pulses. The tunneling conductance d$I$/d$V$ was computed numerically from $I-V$ curves.

\begin{figure}
\centerline{\epsfig{file=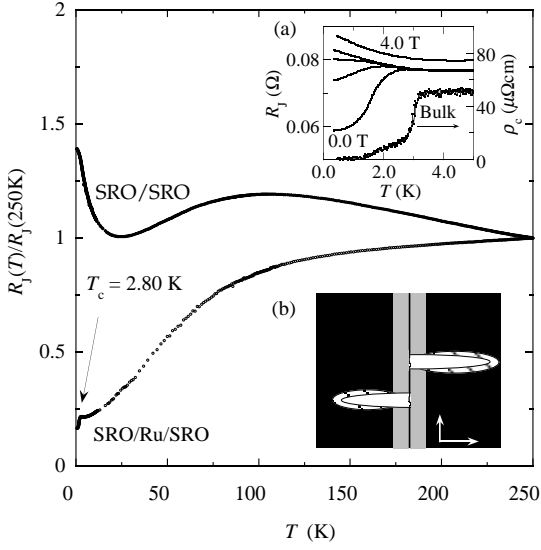,angle=0,width=3.4in}}
\medskip 
\caption{Normalized resistance $R_{\rm J}(T)$ for  SRO/Ru/SRO and SRO/SRO junctions. Insets: (a) $R_{\rm J}(T)$ of the SRO/Ru/SRO junction under fields, from the top to bottom, $B$=4.0, 0.6, 0.4, 0.2 and 0 T, and $\rho_{\rm c}(T)$ of the bulk Ru-containing Sr$_2$RuO$_4$ crystals. (b) Schematic of the SRO/Ru/SRO junction. The gray region is the non-superconducting surface layer of Sr$_2$RuO$_4$ (see text), the white region the Ru inclusions, and shadow region around the Ru inclusions is the 3-K phase.}
\label{1}
\end{figure}

Figure~\ref{1} shows the normalized $R_{\rm J}(T)$ of the SRO/Ru/SRO measured at 1 mA (At this current, the voltage was found to be below 0.08 meV for $T<$3 K.). For comparison, $R_{\rm J}(T)$ of a Ru-free $c$-axis SRO/SRO junction, again measured at 1 mA, is also shown. For the latter, the $R_{\rm J}(T)$ becomes insulating below 25 K, suggesting that the interface must have a significantly high potential barrier.  Since no clear feature associated with superconductivity at the $T_{\rm c}$ of the 1.5-K phase is seen, superconductivity must be strongly suppressed at the surface of Sr$_2$RuO$_4$, resulting in a non-superconducting surface layer (NSSL). The presence of such a NSSL is supported by results obtained on all other junctions we have studied so far, including Pb/Sr$_2$RuO$_4$ \cite{5}, Au/SrTiO$_3$/Sr$_2$RuO$_4$, Au/Sr$_2$RuO$_4$, and In/Sr$_2$RuO$_4$ junctions \cite{18} and STM studies of other group \cite{19}. The d$I$/d$V$ tunneling spectra of SRO/SRO and other normal-metal/SRO junctions show a sharp dip at zero bias, which was found to be unrelated to superconductivity of Sr$_2$RuO$_4$ \cite{18}.

The SRO/Ru/SRO junction, however, exhibits behavior distinctly different from that of the SRO/SRO junction. $R_{\rm J}(T)$ showed a metallic behavior down to 5 K, which is in sharp contrast to the bulk resistivity along the $c$-axis  $\rho_{\rm c}(T)$ of clean Sr$_2$RuO$_4$ \cite{7}. Below 5 K, the $R_{\rm J}(T)$ showed a slight up-turn, which is then followed by a noticeable drop at about 2.80 K. When this drop was suppressed by a magnetic field (inset (a) of Fig.~\ref{1}), the $R_{\rm J}(T)$ showed a insulating behavior at low temperatures, indicating the barrier resistance in this junction was larger than the bulk resistance, at least at low temperatures. Therefore, a clear drop in $R_{\rm J}(T)$ indicate that there must be Ru inclusions at the interface. As will be discussed below, this drop originated from the surface ABSs, which are capable of carrying currents.

The inset (b) of Fig.~\ref{1} shows the schematic of SRO/Ru/SRO junction inferred from above analysis. This picture is also strongly supported by direct examination of the cleaved surfaces of the crystals by an optical microscope, which revealed the presence of scattered Ru inclusions. The Ru inclusions at the cleaved interface were in contact with the NSSL (gray regions) due to unavoidable mismatch between the two cleaved crystals. The overall tunnel barrier around Ru inclusion (thin lines at the interface) must be lower than the rest of the junction where two tunnel barriers from both NSSL (thick lines) are involved, making our tunneling measurements essentially a local probe for areas surrounding the Ru inclusions. Since the contact resistance between the Ru inclusion and the bulk Sr$_2$RuO$_4$ should be smaller in comparison with that between Ru and the NSSL \cite{5}, the voltage drop, which defines the energy scale in the tunneling spectrum, is mainly at the interface between the Ru and NSSL. The Ru inclusions are superconducting by the virtue of the proximity effect \cite{10}. This is actually consistent with the fact that the onset temperature of drop in $R_{\rm J}(T)$ of the SRO/Ru/SRO junction is lower than that in the bulk, as shown in the inset (a) of Fig.~\ref{1}.

\begin{figure}
\centerline{\epsfig{file=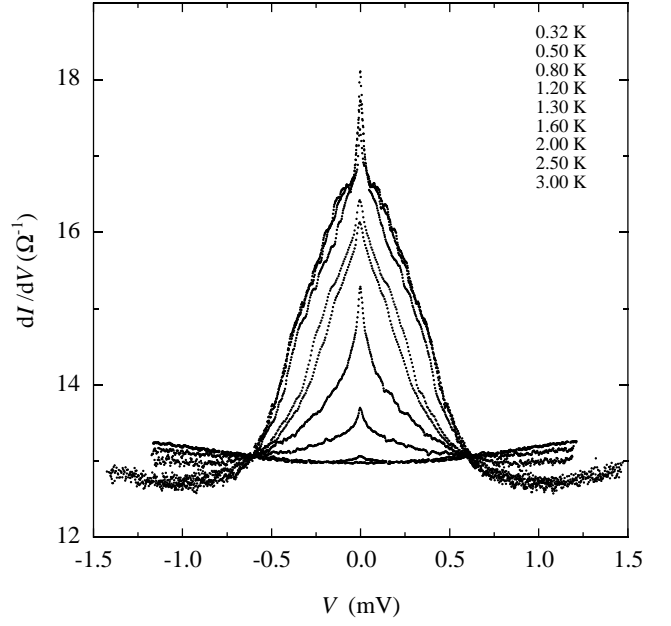,angle=0,width=3.4in}}
\medskip 
\caption{Tunneling spectra d$I$/d$V$ for the SRO/Ru/SRO junction at various temperatures.}
\label{2}
\end{figure}

Figure \ref{2} shows the tunneling spectra of d$I$/d$V$ for the SRO/Ru/SRO junction at various temperatures. A ZBCP occurs below almost the same temperature where the $R_{\rm J}(T)$ shows a drop.  The height of the ZBCP increases with decreasing temperature. Below 1.40 K (the bulk $T_c$), a bell-shaped spectrum with a sharp peak at zero bias emerges.

\begin{figure}
\centerline{\epsfig{file=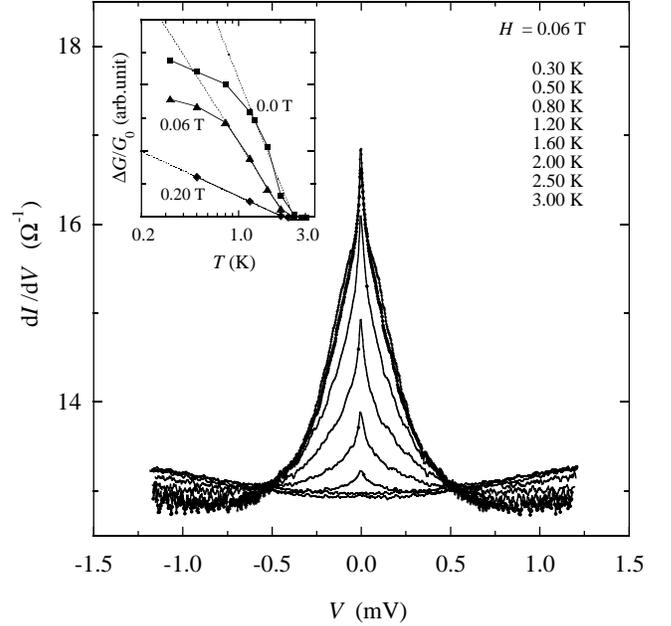,angle=0,width=3.4in}}
\medskip 
\caption{Tunneling spectra d$I$/d$V$ for the SRO/Ru/SRO junction under a fixed field, $H$ = 0.06T, at various temperatures. Inset: temperature (in logarithmic scale) dependence of the normalized height of the ZBCP under different fields. $\Delta G/G_0=({\rm d}I/{\rm d}V|_{V=0}(T) - {\rm d}I/{\rm d}V|_{V=0}(T_c))/{\rm d}I/{\rm d}V|_{V=0}(T_c)$. The dotted line indicates the logarithmic temperature dependence}
\label{3}
\end{figure}

This bell-shaped feature can be suppressed by applying a magnetic field of 0.06 T ($B//c$), as shown in Fig.\ref{3}. This field is approximately the upper critical field along the $c$ axis, $B_{c2}^{//c}$(0), of the primary phase with $T_{\rm c}$ = 1.4 K \cite{20}. Additionally, Fig.~\ref{3} shows that the sharp ZBCP under 0.06 T decreases in height with increasing temperature, disappearing around the $T_{\rm c}$ of the 3-K phase (2.5 K at $B$=0.06 T). This suggests that the sharp ZBCP represents an intrinsic feature of the tunneling spectrum of the 3-K phase.

\begin{figure}
\centerline{\epsfig{file=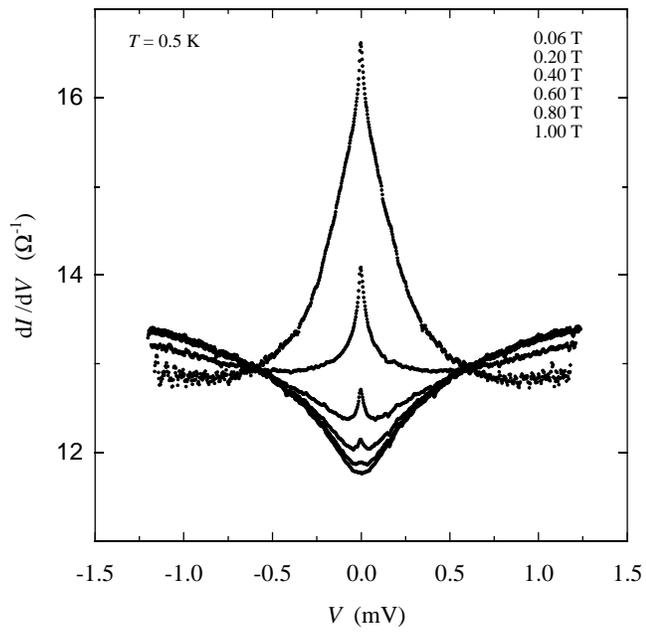,angle=0,width=3.4in}}
\medskip 
\caption{Tunneling spectra d$I$/d$V$ as a function of magnetic field for the SRO/Ru/SRO junction at $T=0.5$ K.}
\label{4}
\end{figure}

The field dependence of the tunneling spectra of the SRO/Ru/SRO junction at 0.5 K is shown in Fig. ~\ref{4}.  The ZBCP is completely suppressed as the field increases up to the $B_{c2}^{//c}$ of the 3-K phase at 0.5 K (=0.9T)\cite{10}. This further demonstrates that the observed sharp ZBCP results from superconductivity of the 3-K phase.  Furthermore, as the ZBCP is fully suppressed, the saturated normal-state background ($B>$0.9 T) resembles the tunneling spectra of Au/SRO \cite{18} and SRO/SRO  as mentioned above, providing further support to our picture for SRO/Ru/SRO junction as shown in Fig. 1.

For the origin of ZBCP, we note that in principle the ABS is not the only explanation.  Several other mechanisms can result in a ZBCP, including: (a) supercurrent leakage; (b) Copper-pair-tunneling caused by proximity effect \cite{21}; (c) Reflectionless tunneling in high-transmittance junctions \cite{22}; and (d) Magnetic and Kondo scattering \cite{23}.   

The possibility of (a) is inconsistent with the large voltage scale involved in the ZBCP we observed. Furthermore, our earlier tunneling measurements on a HgIn/Sr$_2$RuO$_4$ junction revealed a ZBCP similar to that in the cleaved junction when HgIn was driven normal \cite{24}. The mechanisms (b), (c) and (d) are not relevant to our observations as well. For (b), the ZBCP caused by the proximity effect can be suppressed by a field much smaller than critical field. For (c), the ZBCP occurs only in junctions with high transmittance, always accompanied by a gap-like feature, a peak in the spectrum. For (d), the ZBCP is uncorrelated with the occurrence of superconductivity and must show a Zeeman splitting under magnetic fields. These features clearly do not fit into our observations. As a result, we conclude that the ZBCP observed in the SRO/Ru/SRO junction results from the ABS. This is only possible if the 3-K phase in Sr$_2$RuO$_4$ is also a non-$s$-wave superconductor.

The experimental results presented above clearly indicate that the ABSs formed on the surface of Ru inclusions have two distinct spectra depending on whether the bulk Sr$_2$RuO$_4$ is superconducting, which suggests that the 3-K phase has a different pairing symmetry from the 1.5-K bulk phase. This is consistent with the result of recent phenomelogical theory by Sigrist {\it et al} \cite{25}. Below we will discuss the physical origin of these two types of spectrum, and the nature of the crossover. 

The bell-shaped feature seen below 1.4 K (Fig.\ref{2}) is similar to the broad ZBCP observed in Ru-free Sr$_2$RuO$_4$ via point contact \cite{26}. It was suggested that the broad ZBCP originated from the $p$-wave state with a $d$-vector in the form, $d(k)=z(k_x\pm ik_y)$ (fully gapped). Such a $d$-vector will result in ABSs with a continuous spectrum centered at zero energy \cite{27,28}. The average over many sub-gap states will lead to the broadening of the ZBCP. 

The sharp ZBCP is intrinsic to the 3-K phase (see Fig. \ref{3} and \ref{4}). The conductance appears to diverge as the energy approaches zero, suggesting that the ABSs of the 3-K phase are concentrated at zero energy. This is in good agreement with the suggestion that the 3-K phase corresponds to a $p$-wave line-node state \cite{25}, since a $p$-wave pairing with a line node has been shown theoretically to lead to a zero-energy ABS \cite{29}. The $p$-wave states with a line-node can be presented with a $d$-vector in the form of $zk_x$, $zk_y$, or the sum of both. The physical origin of the presence of line node in the 3-K phase can be ascribed to the lower symmetry on the surface of Ru inclusions. On the other hand, it should be mentioned that the possibility of $d$-wave for the 3-K phase can not be excluded completely only from our data, since the $d$-wave superconductors also have a similar zero-energy ABS, leading to a sharp ZBCP \cite{12,13,14,15,16}. 

The crossover of the shape of the ZBCP below the bulk $T_{\rm c}$ may, in principle, be understood in two different ways: (1) The order parameter of the 1.5-K phase with $d(k)=z(k_x\pm ik_y)$ may penetrate into the 3-K phase region and in turn causes a coexistence of order parameters of two phases at $T<1.4$ K in the Ru inclusions through the proximity effect. (2) There is a phase transition \cite{25} associated with a time-reversal symmetry breaking process: the 3-K phase with a line-node transforms into a state with an order parameter close to that of the bulk, $d(k)=z(k_x\pm i\alpha k_y)$ at about bulk $T_{\rm c}$(=1.4 K).

The normalized temperature dependence of the height of the ZBCP ($\Delta G/G_0$) under different fields shown in the inset of Fig. \ref{3} appears to support the phase transition picture. The $\Delta G/G_0$ in zero field can be fitted to a logarithmic $T$-dependence above 1.4 K, below which the $\Delta G/G_0$ increases much more slowly than the logarithmic $T$-dependence. This appears to counter the picture of coexistence of two order parameters in which the $\Delta G/G_0$ should increase faster than the logarithmic $T$-dependence as the second order parameter is added below the $T_{\rm c}$ of the bulk phase. However, this observation is consistent with the phase transition picture. At $B$=0.06 T, the bulk superconductivity of the 1.5-K phase should be suppressed \cite{20}, however $\Delta G/G_0$ still showed a downturn from the logarithmic $T$-dependence below 0.8 K, suggesting that the characteristic field for this crossover is slightly higher than 0.06 T. This is consistent with the fact that a slight broad feature in d$I$/d$V$ is still visible below 0.8 K under 0.06 T (see Fig.~\ref{3}). The $\Delta G/G_0$ under 0.2 T can be fitted well with logarithmic $T$-dependence. The appearance of the central sharp peak below 1.4 K suggests that the crossover associated the phase transition may be incomplete.

Finally, we note that the ZBCP observed in Sr$_2$RuO$_4$ did not split under field as in d-wave superconductors \cite{15,30}. In our experiments, the junction interface formed by cleaving was nominally perpendicular to $c$-axis, with the Andreev reflection occurring in the in-plane direction in order to form an ABS\cite{29}. This is only possible when the surface of the Ru-inclusion was not smooth. Therefore, it is not surprising that a splitting was not found in our spectra. Moreover, for 1.5-K bulk phase, the spectrum of ABS corresponding to $d(k)=z(k_x\pm ik_y)$ is continuous. The field-splitting of individual ABSs will be smeared out \cite{27,28} even for an ideally smooth interface. 

In conclusion, a ZBCP associated with the 3-K superconducting phase in Sr$_2$RuO$_4$ has been observed in cleaved junctions. This ZBCP can be ascribed to surface ABSs, providing the first experimental evidence that the 3-K superconducting phase in Sr$_2$RuO$_4$ is also non-s-wave, possibly a $p$-wave state. In addition, we have observed a crossover in the shape of the ZBCP around bulk $T_{\rm c}$, which has provided a support that there is a phase transition in the 3-K phase region near bulk $T_{\rm c}$.

We would like to acknowledge valuable discussions with M. Sigrist and J. Sauls, and technical assistance from M.A. Rosario.  The work is supported in the US by the NSF through grant DMR-9974327.

\end{document}